\documentclass[12pt]{article}

\usepackage{paper2e}
\usepackage{mydefs2e}

\begin{document}
\renewcommand{\Box}{\Yfund\,}

\newcommand{\ths}{\vartheta}

\newcommand{\N}{$\scr{N}=1$\xspace}

\begin{titlepage}
\preprint{UMD-PP-00-028 \\
JHU-TIPAC-200006} 

\title{Hierarchy Stabilization in\\\medskip
Warped Supersymmetry}

\author{Markus A. Luty}

\address{Department of Physics, University of Maryland\\
College Park, Maryland 20742, USA\\
{\tt mluty@physics.umd.edu}}

\author{Raman Sundrum}

\address{Department of Physics and Astronomy, Johns Hopkins University\\
Baltimore, Maryland 21218, USA\\
{\tt sundrum@pha.jhu.edu}}

\begin{abstract}
We show that exponentially large warp factor hierarchies can be dynamically
generated in supersymmetric compactifications.
The compactification we consider is the supersymmetric extension of the
Randall--Sundrum model.
The crucial issue is the stabilization of the radius modulus for large warp
factor.
The stabilization sector we employ is very simple, consisting of two pure
Yang--Mills sectors, one in the bulk and the other localized on a brane.
The only fine-tuning required in our model is the cancellation of the
cosmological constant, achieved by balancing the stabilization energy against
supersymmetry breaking effects.
Exponentially large warp factors arise naturally, with no very large
or small input parameters.
To perform the analysis, we derive the 4-dimensional effective theory
for the supersymmetric Randall--Sundrum model, with a careful treatment
of the radius modulus.
The manifestly (off-shell) supersymmetric form of this effective lagrangian
allows a straightforward and 
systematic treatment of the non-perturbative dynamics
of the stabilization sector.
%
%
\end{abstract}

\end{titlepage}

\section{Introduction}
Following the work of \Refs{rs1,rs2} there has been a great deal of interest
in the phenomenological possibilities of warped higher-dimensional
spacetimes of the form
\beq
ds^2 = \om^2(y) \eta_{\mu \nu} dx^{\mu} dx^{\nu} + h_{mn}(y) dy^m dy^n,
\eeq
where $x^{\mu}$ ($\mu = 0, \ldots, 3$) are the 4 noncompact spacetime
dimensions, and $y^m$ are compactified.
In particular, the $y$-dependent renormalization of effective four-dimensional
mass scales implied by the `warp factor' $\om(y)$ provides a powerful mechanism
for generating hierarchies in nature.
\Ref{rs1} presented a very simple warped five-dimensional compactification with
an exponential warp factor (the `RS1' model), which exploited this mechanism to
explain the hierarchy between the weak and the Planck scales, without appealing
to supersymmetry.

Warped spacetimes may also be important in models with super\-sym\-metry (SUSY).
One motivation is to allow phenomenological effective field theory 
approaches to
make contact with warped superstring backgrounds \cite{rs1string}.
A particularly interesting string background is
${\rm {\bf AdS}}_5 \times {\bf S}^5$,
which plays a central role in the Maldacena realization of holographic duality
\cite{holography}.
\Refs{rsholo} have emphasized that such dualities may have a profound
connection to the Randall--Sundrum models, based on the (partial)
${\rm {\bf AdS}}_5$ geometry of these models.
Supersymmetry may allow a more precise understanding.
Another example is eleven-dimensional heterotic M-theory compactified on a
six-dimensional Calabi--Yau space and an ${\bf S}^1/{\bf Z}_2$ orbifold
\cite{hw}.
This \N supersymmetric theory has been taken as the starting point for
phenomenological studies, where the warp factor may play an important role
\cite{hwphen}.
There are also purely phenomenological motivations;
warp factors can generate hierarchies required in realistic supersymmetric
theories \cite{warpphen}.
It is also an interesting open question to ask what patterns of SUSY
breaking can arise in warped spacetimes.
In the future we hope to focus on the effect of warping on
higher-dimensional SUSY mediation mechanisms
such as anomaly-mediated SUSY breaking \cite{amsb},
gaugino-mediated SUSY breaking \cite{gmsb},
and radion-mediated SUSY breaking \cite{rmsb}.

In this paper we will study the minimal supersymmetric extension of 
the simplest
warped compactification, namely  RS1.
This extension has been constructed in \Refs{susyrs1}.
Our first result is a derivation of the 4-dimensional effective theory of
the supersymmetric RS1 model valid at long wavelengths, including a careful
treatment of the radius modulus.%
\footnote{A related derivation and discussion of the four-dimensional 
effective theory by J. Bagger, D. Nemeschansky and Ren-Jie Zhang
will appear at the same time as the present paper.}
This effective lagrangian is valid to 2-derivative order,
but to all orders in the fields, including the radion field.
The effective lagrangian will be presented in terms of off-shell SUSY multiplets,
which will greatly simplify the analysis of non-perturbative effects and
SUSY breaking.

The other main result of our paper is a dynamical mechanism to stabilize the
radius modulus in the supersymmetric RS1 model.
This mechanism naturally stabilizes the radius at a sufficiently large
value that the warp factor hierarchy across the extra dimension is large.
The stabilization sector consists of two super-Yang--Mills (SYM) sectors,
one in the bulk and the other localized on one of the 4-dimensional
boundaries.
The radius of the extra dimension is stabilized by the balance between
brane and boundary gaugino condensate contributions to the supergravity
(SUGRA) potential.
We first proposed this mechanism in \Ref{ls}, where it was shown to stabilize
the radius in a supersymmetric compactification with negligible warp factor.
We stress that for any value of the warp factor, the mechanism is completely
natural (except for the cosmological constant problem) and controlled in an
effective field theory expansion.
In the non-supersymmetric RS1 model, a simple classical mechanism that
stabilized a large warp factor was presented in \Ref{gw}.
The supersymmetric mechanism we present here is intrinsically non-perturbative.

We believe that it is an important development to have a supersymmetric model
of radius stabilization that is both \emph{complete} and \emph{calculable\/}.
Moduli describing the size and shape of the extra dimensions are a generic
feature of higher-dimensional compactifications with supersymmetry,
and in particular superstring theory.
These moduli must be stabilized both to avoid phenomenological and cosmological
problems of light scalars, and also to select an appropriate vacuum.
This problem has been extensively discussed in string-inspired contexts;
see \eg\ \Ref{dine}.
The stabilization problem is especially severe because of the constraints
of higher-dimensional local supersymmetry.
Our model gives a simple stabilization mechanism consistent with these
constraints, even if it does not display the full complexity of string
compactifications.
We hope that some of the tools we have developed can be extended to
superstring/M-theory.

This paper is organized as follows.
In section 2 we describe the model we will study.
In section 3 we derive the supersymmetric 4-dimensional effective field theory
of the supersymmetric RS1 model.
In section 4 we analyze the non-perturbative gauge dynamics needed for
stabilization using the effective 4-dimensional description.
These results are summarized and discussed in Section 5.
In the interest of readability, some details of the derivation of the
effective theory in section 3 are relegated to the appendix, which however
gives a self-contained account.

\section{The Model}
The theory we are interested in is minimal 5-dimensional SUGRA, where the
$5^{\rm th}$ dimension is a finite interval realized as a
${\bf S}^1 / {\bf Z}_2$ orbifold.
We will also couple this theory to matter and gauge fields in the bulk or
localized on the orbifold boundaries.

Our starting point is the on-shell lagrangian for 5-dimensional SUGRA
\cite{sugra5}
\beq\eql{SUGRA5}
\bal
\scr{L}_{\rm SUGRA,5} = -M_5^3 & \biggl\{ \sqrt{-G} \left[
\sfrac{1}{2} R(G)
+ \sfrac{1}{4} C^{MN} C_{MN} - 6 k^2 \right]
\\
& \quad
+ \frac{1}{6\sqrt{6}} \ep^{MNPQR} B_M C_{NP} C_{QR}
+ \hbox{\rm fermion\ terms} \biggr\},
\eal\eeq
where $M, N, \ldots = 0, \ldots, 3, 5,$ are 5-dimensional
spacetime indices, $G_{MN}$ is the 5-dimensional metric,
$C_{MN} = \partial_M B_N - \partial_M B_N$
is the field strength for the graviphoton $B_M$,
and $k$ is a mass scaled defined so that
$-6 M_4^3 k^2$ is the 5-dimensional cosmological constant.
Unbroken SUSY requires that the cosmological constant
have {\bf AdS} sign ($k^2 > 0$).
In order to realize this theory on an ${\bf S}^1 / {\bf Z}_2$ orbifold,
the ${\bf Z}_2$ parity assignments of the bosonic fields must be
taken as in Table~1.

We now couple 5-dimensional SUGRA to localized energy density
on the orbifold boundaries:
\beq[bdy]
\De\scr{L}_5 = - \de(\ths) \sqrt{-g_1}\, V_1
- \de(\ths - \pi) \sqrt{-g_2}\, V_2,
\eeq
where $g_{1,2}$ are the induced 4-dimensional metrics on the boundaries,
and $V_{1,2}$ are constants, and $-\pi < \ths \le \pi$ parameterizes
the 5$^{\rm th}$ dimension.
This theory admits the Randall--Sundrum solution \cite{rs1}
\beq\bal
\eql{RSsoln}
ds^2 &= e^{-2 k r_0 |\ths|} \eta_{\mu\nu} dx^\mu dx^\nu
+ r_0^2 d\ths^2,
\\
& \qquad
B_\ths = b_0,
\qquad
B_\mu = 0,
\eal\eeq
provided that
\beq
k = \frac{V_1}{6 M_5^3} = -\frac{V_2}{6 M_5^3}.
\eeq
This metric is a slice of {\bf AdS}$_5$.
The exponential factor $e^{-2 k r_0 |\vth|}$ is the `warp factor' that gives
rise to mass hierarchies across the $5^{\rm th}$ dimension.
The theory including the boundary terms \Eq{bdy} can be made
supersymmetric by the addition of suitable fermion terms, and
the `vacuum' solution \Eq{RSsoln} then preserves 4 real supercharges
\cite{susyrs1}.
The bulk lagrangian \Eq{SUGRA5} is invariant under 8 real supercharges,
but half of the supersymmetry is explicitly broken by the
orbifold projection and the boundary terms.

\Eq{RSsoln} is a solution for any value of $r_0$ and $b_0$;
$r_0$ is the radius of the compact ${\bf S}^1$, while $b_0$ is the
Aharonov-Bohm phase of the graviphoton around the ${\bf S}^1$.
When we consider fluctuations about the solution \Eq{RSsoln}, these integration
constants become propagating massless modes.
The mode corresponding to $r_0$ (the radion) is particularly important,
since it controls the couplings in the 4-dimensional effective theory.
In this paper we will show how to stabilize the radion in the interesting
case where the warp factor is a large effect.

\begin{table}[t] 
\centering
\begin{tabular}{c|c}
Field & ${\bf Z}_2$ Parity \\
\hline
$G_{\mu\nu}$ & $+$ \\
$G_{5\mu}$ & $-$ \\
$G_{55}$ & $+$ \\
$B_\mu$ & $-$ \\
$B_5$ & $+$ \\
\end{tabular}
\\
\capt{Bosonic fields of 5-dimensional SUGRA with
their ${\bf Z}_2$ parity assignments.}
\end{table}

In addition, we will couple this theory to bulk super-Yang--Mills (SYM)
theory.
The minimal 5-dimensional SYM multiplet consists of a vector field
$A_M$, a real scalar $\Phi$,
and a symplectic Majorana gaugino $\la^j$ ($j = 1, 2$).
The bulk lagrangian is \cite{gun}
\beq\bal
\scr{L}_5 &= -\sqrt{-G}\, \frac{1}{2 g_5^2} \tr F^{MN} F_{MN}
- \frac{1}{2 \sqrt{6}\, g_5^2} \ep^{MNPQR} B_M \tr F_{NP} F_{QR}
\\
&\qquad\qquad
+ \hbox{\rm scalar\ and\ gaugino\ terms}.
\eal\eeq
The SYM fields are taken to transform under the orbifold ${\bf Z}_2$
as shown in Table 2.
The even fields form an \N SYM multiplet.

To obtain realistic models we will couple
these bulk fields to fields localized on the orbifold boundaries.
Working out these couplings and verifying that they preserve supersymmetry
is nontrivial.
An off-shell construction of the boundary couplings was given by \Ref{zucker}
using the method of Mirabelli and Peskin \cite{peskin}.
The off-shell couplings of bulk SYM to SUGRA were constructed in 
\Ref{symsugra}.
It is clearly crucial for the results of this paper that these couplings exist
and preserve SUSY.
However, the results of this paper will be derived using only the on-shell
bosonic lagrangian together with consistency arguments.

\begin{table}[t] 
\centering
\begin{tabular}{c|c}
Field & ${\bf Z}_2$ Parity \\
\hline
$A_\mu$ & $+$ \\
$A_5$ & $-$ \\
$\Phi$ & $-$ \\
$\la^1$ & $+$ \\
$\la^2$ & $-$ \\
\end{tabular}
\\
\capt{Fields of 5-dimensional super-Yang--Mills sector with
their ${\bf Z}_2$ parity assignments.}
\end{table}

We can now summarize the
theory that we will analyze in this paper.
The theory consists of minimal 5-dimensional SUGRA, with a SYM sector
in the bulk, and an additional SYM sector on one of the orbifold
boundaries, the `hidden brane.'
The bulk lagrangian has dimensionful parameters $M_5$ and $g_5$ that we
take to be of order the Planck scale.
Additionally, we assume that there is a SUSY breaking sector also localized on
the hidden brane.
The SYM multiplets together with the SUSY breaking sector
will play the role of the radius stabilization sector,
as we will see.
For a fully realistic model, one would want to add standard model fields,
presumably some or all of them localized on the other boundary, the
`visible brane.'
These play no role in the stabilization dynamics.
We will study complete realistic models in future work.

\section{The 4-Dimensional Effective Lagrangian}
At sufficiently low energies, the dynamics of the theory above is
approximately 4-dimensional.
The matching scale between the 5-dimensional and 4-dimensional effective
theories is determined by the mass of the lowest KK mode, given by 
\cite{rs1,rs2}
\beq[matchscale]
m_{\rm KK}^2 \sim \left( \frac{k}{1 + e^{\pi k r_0}} \right)^2.
\eeq
We assume that the theory is weakly interacting at this scale,
justifying the use of classical matching.
This will be true as long as the radius of compactification $r_0$
and the radius of curvature $1/k$ are larger than the 5-dimensional
Planck length.

In this section, we will derive the 4-dimensional effective theory below
the scale \Eq{matchscale}.
Our strategy is to match enough bosonic terms between the 5-dimensional and
4-dimensional lagrangians, so that we can infer the remaining terms using \N
SUSY.
The justification of some of the steps is relegated to an appendix.
The appendix gives a complete self-contained derivation,
including a discussion of some subtleties of classical matching.

We begin by considering the massless bosonic fields arising from the
5-dimensional SUGRA sector.
The solution \Eq{RSsoln} has undetermined
integration constants $r_0$ and $b_0$
whose long-wavelength fluctuations are massless moduli.
Also, unbroken 4-dimensional Lorentz invariance implies
that there is a massless graviton in the 4-dimensional effective
theory.
These massless 4-dimensional fluctuations can be parameterized by
making the replacements
$\eta_{\mu\nu} \to g_{\mu\nu}(x)$, $r_0 \to r(x)$, and
$b_0 \to b(x)$ in \Eq{RSsoln}:
\beq[rmetric]
\bal
ds^2 &= e^{-2 k r(x) |\ths|} g_{\mu\nu}(x) dx^\mu dx^\nu
+ r^2(x) d\ths^2,
\qquad
-\pi < \ths \le \pi,
\\
& \qquad B_{\ths}(x, \ths) = b(x),
\qquad
B_\mu(x, \ths) = 0.
\eal
\eeq
If this satisfied the 5-dimensional equation of motion,
one could obtain the classical 4-dimensional effective action
by substituting \Eq{rmetric} into the 5-dimensional action and
integrating over the $5^{\rm th}$ dimension.
\Eq{rmetric} does not satisfy the 5-dimensional
equations of motion \cite{cgr}.
However, in the appendix we show that for the metric in \Eq{rmetric},
this `\naive' procedure gives a result that differs from the
exact classical effective action only by terms with four or
more $x$ derivatives.
We can therefore use the metric in \Eq{rmetric} to parameterize the radion at
leading order in the derivative expansion.%
\footnote{\Ref{cgr} gives an alternate parameterization of the radion
that satisfies the 5-dimensional equations of motion at linear
order in fluctuation fields, but to all orders in $x$ derivatives.}
For the graviphoton, the \naive procedure does not work;
the graviphoton can still be parameterized by $b(x)$ defined by \Eq{rmetric},
but there is a nontrivial correction to the classical effective lagrangian that
is computed in the appendix.
However, to determine the effective theory it is sufficient to know the terms
that depend only on the radion, which can be determined by substituting
\Eq{rmetric} into the 5-dimensional action.
The terms depending on the graviphoton can then be inferred from SUSY.
Therefore, the calculation of the graviphoton effective lagrangian
carried out in the appendix serves only as a redundant check on our
results.

We turn to the 5-dimensional SYM sector.
It is straightforward to verify that
\beq[gaugezero]
A_\mu(x, \ths) = a_\mu(x),
\qquad
A_\ths(x, \ths) = 0
\eeq
is a solution to the 5-dimensional equations of motion if
$a_\mu(x)$ is a solution to the 4-dimensional YM equation of
motion.
Therefore, $a_\mu(x)$ parameterizes a 4-dimensional vector
zero mode.
The fact that the zero mode is independent of $\ths$ despite the
presence of the warp factor can be traced to the classical conformal invariance
of 4-dimensional Yang--Mills theory.
Note that there are no massless $A_\ths$ or $\Phi$ fluctuations because of the
orbifold projection.

We wish to relate the massless bosonic fields defined above (and 
their fermionic
superpartners) to a manifestly \N supersymmetric formulation of the
4-dimensional effective theory.
The massless bosonic fields are two real scalars $r(x)$ and $b(x)$,
a real vector multiplet $a_\mu(x)$, and the metric $g_{\mu\nu}(x)$.
Given that these bosonic fluctuations are part of an \N locally supersymmetric
theory, they can be parameterized by one chiral superfield $T$, one
non-Abelian vector superfield $V$, and the minimal SUGRA multiplet.
The most general effective lagrangian at 2-derivative order can be written
\beq[SUGRAgeneral]
\scr{L}_{\rm 4,eff} = \myint d^4\th\, \phi^\dagger \phi\, f(T, T^\dagger)
+ \left[ \myint d^2\th\, S(T) \tr(W^\al W_\al) + \hc \right].
\eeq
There is no superpotential for $T$ because the radion modulus
does not have a potential.
We are using the superconformal approach to SUGRA \cite{superconf}.
The field $\phi$ is the superconformal compensator \cite{gates,superconf} that
is responsible for breaking the local superconformal symmetry down to local
super-Poincar\'e:
\beq[scgauge]
\phi = 1 + \th^2 F_\phi.
\eeq
$F_\phi$ is the scalar auxiliary field of the minimal off-shell \N SUGRA
multiplet.
We are using superspace notation as a shorthand for expressions that can
be rigorously defined using the superconformal tensor calculus approach
\cite{superconf}.
In particular, factors of the metric (or vierbein) are implicit in this
notation.

We now make a holomorphic field redefinition $S(T) \to T / g_5^2$
in the effective theory so that the effective lagrangian has the form
\beq[SUGRAgeneral2]
\scr{L}_{\rm 4,eff} = \myint d^4\th\, \phi^\dagger \phi\, f(T, T^\dagger)
+ \left[ \myint d^2\th\, \frac{T}{g_5^2} \tr(W^\al W_\al) + \hc \right].
\eeq
 From this, we have
\beq
\frac{1}{2 g_4^2} = \frac{\Re(T)}{g_5^2}.
\eeq
We can also calculate the 4-dimensional gauge coupling $g_4$ by substituting
the zero mode gauge field \Eq{gaugezero} into the 5-dimensional action
and integrating over the $5^{\rm th}$ dimension.
This yields
\beq
\frac{1}{g_4^2} =
\frac{2\pi r}{g_5^2}.
\eeq
Therefore, we see that
\beq[ReT]
\Re(T) = \pi r.
\eeq

Similarly, from \Eq{SUGRAgeneral2} we see that $\Im(T)$ is proportional to
the 4-dimensional theta angle, which in turn is proportional to $B_\ths$
from the mixed Chern-Simons term in the 5-dimensional theory:
\beq
\De\scr{L}_5 &= -\frac{1}{2\sqrt{6}\, g_5^2}
\ep^{MNPQR} B_M \tr \left( F_{NP} F_{QR} \right)
\nonumber\\
&=  -\frac{1}{2\sqrt{6}\, g_5^2} \ep^{\mu\nu\rho\si} B_{\ths}
\tr\left( F_{\mu\nu} F_{\rho\si} \right) + \cdots
\eeq
This determines
\beq[ImT]
\Im(T) = \frac{2\pi}{\sqrt{6}}\, b.
\eeq
Thus we have fixed the relation between $T$ and the component fields
$r(x)$ and $b(x)$.
Note that \Eqs{ReT} and \eq{ImT} are exactly the same as in flat
space \cite{ls}.
This is ultimately due to the classical conformal invariance of Yang--Mills
theory in 4 dimensions.

It still remains to fix the relation between the metric that appears in
the 4-dimensional \N SUGRA multiplet, and the metric $g_{\mu\nu}$ defined
by \Eq{rmetric}.
This is nontrivial because in the 4-dimensional effective theory, we have the
freedom to make field redefinitions $g'_{\mu\nu} = c(r) g_{\mu\nu}$, where
$c(r)$ is an arbitrary function.
However, such field redefinitions in general do not preserve the property that
$T$ transforms independently of the 4-dimensional SUGRA multiplet
under \N SUSY.
In the appendix, it is shown that imposing this condition implies that
the two metrics are identical (as implicitly
assumed in the notation used above).

Expanding the 4-dimensional SUGRA lagrangian \Eq{SUGRAgeneral} in component
fields, we obtain
\beq[genSUGRA4]
\bal
\scr{L}_{\rm SUGRA,4} = \sqrt{-g} \biggl[ &
-\frac{1}{6} f R(g)
- \frac{1}{4 f} (f_T \partial^\mu T - \hc) (f_T \partial_\mu T - \hc)
\\
&
- f_{T^\dagger T} \partial^\mu T^\dagger \partial_\mu T
+ \hbox{\rm fermion\ terms} \biggr],
\eal\eeq
where $f_T = \partial f / \partial T$, {\it etc\/}.,
and $R(g)$ is the 4-dimensional Ricci scalar associated with the metric $g$.
As discussed above, the terms depending on the metric and the radion $r$ can
be obtained by substituting \Eq{rmetric} into the 5-dimensional
action and integrating over the $5^{\rm th}$ dimension.
We can use this procedure to determine $f$ by calculating the coefficient
of $R(g)$ (see \Eq{genSUGRA4}).
One obtains
\beq
f = \frac{3 M_5^3}{k} \left( e^{-2\pi k r} - 1 \right).
\eeq
Note that for $r \to r_0$ $f$ is the 4-dimensional Planck scale computed in
\Ref{rs1}.
The graviphoton Aharonov--Bohm phase cannot contribute to the coefficient
of $R(g)$.
Using \Eqs{ReT} and \eq{ImT} therefore gives
\beq
f(T, T^\dagger) = +\frac{3 M_5^3}{k} \left( e^{-k(T + T^\dagger)} - 1 \right).
\eeq
Having fixed the function $f$, the other 2-derivative terms in \Eq{genSUGRA4}
that depend on $r$ and $b$ are fixed.
In the Appendix we show that these agree with a direct matching calculation,
giving a highly nontrivial check of this derivation.

We now turn to fields localized on the boundary.
Note that in terms of the components fields, we have chosen coordinates so
that the warp factor is unity at $\ths = 0$ (the hidden brane).
Therefore the radion (as parameterized above) does not couple to the
fields on the hidden brane.
Correspondingly, it is shown in the appendix that the terms arising from
the hidden brane are independent of $T$.
Therefore the general form of the effective lagrangian involving the
hidden fields is
\beq[Lhid]
\bal
\scr{L}_{\rm 4,hid} &= \myint d^4\th\, \phi^\dagger \phi\,
f_{\rm hid}(\Si, \Si^\dagger)
\\
&\qquad + \myint d^2\th \left[ S_{\rm hid}(\Si) \tr W'^\al W'_\al
+ \phi^3 {\cal W}_{\rm hid}(\Si) \right]
+ \hc,
\eal
\eeq
where $\Si$ are hidden sector chiral multiplets and $W'_\al$ is the field
strength of the hidden sector gauge multiplets.
The terms arising from the visible brane do have couplings to the radion,
since by \Eq{rmetric} the induced metric on the brane is
$e^{-2 \pi k r(x)} g_{\mu\nu}(x)$.
The unique supersymmetrization of these terms is
\beq[Lvis]
\bal
\scr{L}_{\rm 4,vis} &= \myint d^4\th\, \phi^\dagger \phi\,
e^{-k(T + T^\dagger)}
f_{\rm vis}(Q, Q^\dagger) 
\\
&\qquad + \myint d^2\th \left[ S_{\rm vis}(Q) \tr \tilde{W}^\al \tilde{W}_\al
+ \phi^3 e^{-3 k T} {\cal W}_{\rm vis}(Q) \right] 
+ \hc,
\eal
\eeq
where $Q$ is a visible sector chiral multiplet, and $\tilde{W}_\al$ is the
field strength of the visible sector gauge multiplets.
Note that \Eq{Lvis} has the same form as \Eq{Lhid} with
$\phi$ replaced by $\phi e^{-k T}$.
This is not a coincidence.
The radion dependence of $\scr{L}_{\rm 4,vis}$ is entirely due to the
fact that the induced metric is a Weyl rescaling of $g_{\mu\nu}$,
which is precisely equivalent to a rescaling of the conformal compensator
$\phi$.

Comparing \Eqs{Lhid} and \eq{Lvis} one readily sees that, relative to
fundamental mass parameters, physical mass scales in the visible sector
(including UV regulator and renormalization scales) are rescaled by
a factor of $e^{-k\pi r_0}$, while scales on the hidden sector are
not.
For $k > 0$, mass scales are suppressed on the visible sector, while for
$k < 0$ mass scales are enhanced on the visible sector.
This is the warp factor effect that can naturally generate exponentially
large hierarchies.

It is more conventional to describe the kinetic terms in supergravity
in terms of the \Kahler potential.
This is given by
\beq
K \equiv -3 M_4^2 \ln\left[
- \frac{f(T, T^\dagger) + f_{\rm vis}(Q, Q^\dagger)
+ f_{\rm hid}(\Si, \Si^\dagger)}{3 M_4^2} \right].
\eeq
The properties of supersymmetry breaking and renormalization are easier
to see in terms of $f$, but the \Kahler potential is more useful for
determining the sigma model couplings of the bosonic fields.

\section{The Radius Modulus Effective Potential}
In this section, we construct the effective potential for the model
described above and minimize the potential to show that the the radius is
in fact stabilized.
The model was analyzed in \Ref{ls} for the case where the warp factor
is a small effect, $e^{-kT} \simeq 1$.
We will therefore be interested in the case where the warp factor is a
large effect.

Just below the KK matching scale \Eq{matchscale}, the 4-dimensional 
effective theory
is
\beq[startL]
\bal
\scr{L}_{\rm 4,eff} &=\frac{3 M_5^3}{k} \myint d^4\th\, \phi^\dagger \phi
\left( e^{-k(T + T^\dagger)} - 1 \right)
\\
&\qquad + \myint d^2\th \left(
\frac{T}{g_5^3} \tr W^\al W_\al + \frac{1}{2 g_4^2} \tr W'^\al W'_\al \right)
+ \hc
\\
&\qquad + \scr{L}_{\rm SB}.
\eal\eeq
Here the first gauge kinetic term arises from the bulk SYM sector,
while the second arises from the SYM sector localized on the hidden brane.
$\scr{L}_{\rm SB}$ is the lagrangian for the SUSY breaking sector,
also assumed to be localized on the hidden brane.
We are using coordinates where the warp factor is unity on the hidden brane (so
that $\scr{L}_{\rm SB}$ is independent of $T$).
There are therefore two cases to consider: the warp factor at the
visible brane is either smaller or larger than unity.
In the formulas above, these cases correspond to $k > 0$ and $k < 0$,
respectively, so we can analyze both cases using \Eq{startL}.
Classical matching is justified by assuming that the asymptotically free gauge
forces are weak at the KK matching scale, and that the spacetime curvature is
also small, $|k| < M_5$.

The SYM sectors become strong in the infrared of the 4-dimensional
effective theory and give rise to a dynamical superpotential
from gaugino condensation.
In addition, the hidden SUSY breaking sector is assumed to dynamically
generate a nonzero vacuum energy.
This vacuum energy will be positive provided that SUGRA is a perturbation
to the SUSY breaking dynamics.
We also assume that the SUSY breaking dynamics has a mass gap,
except for the Goldstino.
The effective lagrangian below the scale where these effects occur is then
\beq[Leff4]
\bal
\scr{L}_{4,\rm eff} &= \frac{3 M_5^3}{k}\myint d^4\th\,
\phi^\dagger \phi
\left( e^{-k(T + T^\dagger)} - 1 \right)
+ \left[ \myint d^2\th \phi^3 \left(
a e^{-\zeta T} + c \right) + \hc \right]
\\
&\qquad - V_{\rm SB} + \hbox{\rm Goldstino\ terms}.
\eal\eeq
If the bulk SYM gauge group is $SU(N)$, we have
\beq[abdef]
a = \scr{O}\left( \frac{1}{ N^4 g_5^6} \right),
\qquad
\zeta = \frac{16\pi^2}{3 N g_5^2}.
\eeq
The exact $T$ dependence in the superpotential term of \Eq{Leff4} is fixed by
holomorphy and the anomalous shift symmetry in $T$ \cite{ls}.

It is straightforward to integrate out the auxiliary fields for $T$ and
$\phi$ to obtain the effective potential.
However, additional insight into the form of the answer is given by writing
the lagrangian in terms of the `warp factor superfield'
\beq
\om \equiv \phi e^{-k T}
\eeq
in place of $T$.
This gives
\beq[Lsimp]\bal
\scr{L}_{4,\rm eff} &= \frac{3 M_5^3}{k}\myint d^4\th\,
\left( \om^\dagger \om - \phi^\dagger \phi \right)
+ \left[ \myint d^2\th \left(
a \phi^{3 - n} \om^n  + c \phi^3 \right) + \hc \right]
\\
&\qquad - V_{\rm SB} + \hbox{\rm Goldstino\ terms},
\eal\eeq
where
\beq
n \equiv \frac{\zeta}{k}.
\eeq
 From \Eq{Lsimp} one can immediately read off the potential
\beq
\!\!\!\!\!\!\!\!
V_{\rm eff} &= \frac{k}{3 M_5^3} \left(
n^2 |a|^2 (\om^\dagger \om)^{n - 1} - |(3 - n) a \om^n + 3c |^2 \right)
+ V_{\rm SB}
\\\nonumber
&= \frac{k}{3 M_5^3} \left[
n^2 |a|^2 |\om|^{2(n - 1)} - (n - 3)^2 |a|^2 |\om|^{2n}
- 9 |c|^2
- 6(n - 3) |a| |c| |\om|^n \cos\ga \right]
\\
\eql{Vtemp}
&\qquad + V_{\rm SB},
\eeq
where
\beq
\ga \equiv
\mathop{\rm arg}(a) - \mathop{\rm arg}(c) + n \mathop{\rm arg}(\om).
\eeq
We now minimize the potential as a function of $|\om|$ and $\ga$.

%

We first consider $k > 0$, corresponding to the case where
the warp factor is smaller than unity on the visible brane.
If the warp factor is an important effect, then $|\om| \ll 1$ and we can
neglect the second term in \Eq{Vtemp} compared to the first.
(We assume that $n$ is not much larger than unity.)
Minimizing with respect to $\ga$ simply sets $\cos\ga = \sgn(n - 3)$.
There is a nontrivial minimum provided that $n > 3$, which is satisfied provided
that the bulk SYM sector is weakly coupled at the KK matching scale.
We then obtain
\beq[soln1]
|\om| = e^{-\pi k r} = \left( \frac{3 (n - 3)}{n (n - 1)}
\, \frac{|c|}{|a|} \right)^{1/(n - 2)}.
\eeq
We see that for any given $n$ we can obtain $|\om| \ll 1$ provided
that $|c| / |a|$ is sufficiently small.%
\footnote{This assumes that $n$ is not too large.
The regime $n \gg 1$ corresponds to $k \ll \zeta$, \ie small 
bulk curvature.
As shown in \Ref{ls}, the model also stabilizes the radius in this regime.}
This is perfectly natural, since $|c|$ is exponentially
small in terms of the fundamental couplings.
Thus, if we want to use the small warp factor to explain some mass
hierarchy in nature, the small warp factor itself can be explained in terms
of order-1 fundamental parameters in this model of stabilization.

To complete our analysis of this case, we find the other extrema of 
the potential.
There is an obvious extremum where $|\om| \to 0$.
It is easy to check that this has higher energy than the solution \Eq{soln1}.
We must also look for solutions with $|\om| \sim 1$.
In this case we can neglect the last term of \Eq{Vtemp} since $|c| \ll |a|$.
This gives another extremum
\beq[wrong]
|\om| = \left(\frac{n (n - 1)}{(n - 3)^2} \right)^{1/2}.
\eeq
However this solution has $|\om| > 1$ (which is outside the physical region
$r > 0$) for $n > 1$ and is therefore unphysical for the values of $n$ we are
considering. It is also easy to see that this extremum has higher 
energy than the
solution \Eq{soln1}.

Combining the results above,
we conclude that \Eq{soln1} is in fact the true (global) minimum.
In order to cancel the 4-dimensional cosmological constant, we note that
the term $-9 |c|^2$ in \Eq{Vtemp} dominates the vacuum energy in the
solution, so we must fine-tune
\beq
V_{\rm SB} \simeq \frac{3 |c|^2}{M_4^2},
\eeq
where $M_4^2 = M_5^3 / k$.
Note that this gives $V_{\rm SB} > 0$, as desired.
We obtain
\beq
m_{3/2}^2 
= \frac{|c|^2}{M_4^4} \sim \frac{V_{\rm SB}}{M_4^2}.
\eeq
The masses of the radion fields at the minimum of the potential
is straightforward to work out using the component
lagrangian given above, or in terms of the standard 4-dimensional
supergravity potential.
Parameterizing the radion by $\om$ greatly simplifies the calculation.
We find
\beq
m^2_{\rm scalar} = \frac{|c|^2}{M_4^4}\,
\frac{(n - 2) (n - 3)^2}{n - 1},
\qquad
m^2_{\rm pseudoscalar} = \frac{|c|^2}{M_4^4}\,
\frac{n (n - 3)^2}{n - 1}.
\eeq
Note that $m_{\rm scalar} \sim m_{\rm pseudoscalar} \sim m_{3/2}$.
The radion is lighter than the KK matching scale \Eq{matchscale}
provided that $|c|/M_5^3 \ll |\om|$, which is guaranteed by
\Eq{soln1} since $|c| \ll |a| \ll M_5^3$.

We now consider $k < 0$, corresponding to the case that the warp factor
is larger than unity on the visible brane.
Note that in this case $n < 0$.
We again look for solutions where the warp factor is a large effect,
so that $|\om| \gg 1$.
We can therefore neglect the first term of \Eq{Vtemp} compared
to the second.
Because the factor in front of \Eq{Vtemp} is now negative, minimizing
with respect to the phase $\ga$ now gives $\cos\ga = -\sgn(n - 3)$.
We then obtain the solution
\beq[soln2]
|\om| = \left( \frac{|n - 3|}{3} \, \frac{|a|}{|c|}
\right)^{1/|n|}.
\eeq
We see that $|\om| \gg 1$ provided that $|c| \ll |a|$.
Again, \Eq{wrong} is an extremum, as is $|\om| \to +\infty$.
As before, \Eq{wrong} is outside the physical region, and both
\Eq{wrong} and the `runaway' solution $|\om| \to +\infty$ have higher
energy than the solution \Eq{soln2}.

Together, these results imply that \Eq{soln2} is in fact the true (global)
minimum.
In order to cancel the 4-dimensional cosmological constant, we note that
the first term in \Eq{Vtemp} dominates the vacuum energy, and we must
fine-tune
\beq
V_{\rm SB} \simeq \frac{3 |c|^2}{M_4^2} \, \frac{n^2}{(n - 3)^2},
\eeq
where $M_4^2 = M_5^3 |\om|^2 / |k|$.
Again $V_{\rm SB} > 0$ as desired.
We find
\beq
m_{3/2}^2 = \frac{|c|^2}{M_4^4}\, \frac{(n - 6)^2}{(n - 3)^2}
\sim \frac{V_{\rm SB}}{M_4^2}.
\eeq
The radion masses are
\beq
m^2_{\rm scalar} = m^2_{\rm pseudoscalar}
= \frac{|c|^2}{M_4^4}\, |n|^2 |\om|^4.
\eeq
Note that $m_{\rm scalar} = m_{\rm pseudoscalar} \gg m_{3/2}$
in this case.
The radion is lighter than the KK matching scale \Eq{matchscale}
provided only that $|c|/M_5^3 \ll 1$.

We conclude that the simple model we are considering does in fact
stabilize the radius modulus in the regime where the warp factor is
large, provided only that $|c| \ll |a|$.
This works both in the case where the warp factor is largest on the
hidden brane or on the visible brane.
In both cases, the cosmological constant can be cancelled by positive
vacuum energy from the SUSY breaking sector.

\section{Discussion}

Let us summarize what has been accomplished.
The 4-dimensional effective lagrangian describing the supersymmetric
Randall--Sundrum model at long distances was derived.
Like the non-supersymmetric Randall--Sundrum model it has a vanishing potential
for the radius modulus, now a chiral superfield.
We also showed that the mechanism proposed in \Ref{ls} stabilizes this modulus
in the interesting regime where the warp factor is a large effect.

The stabilizing sector consists of two types of supersymmetric Yang-Mills
sectors, one in the bulk and the other on one of the boundaries, the `hidden
brane.'
These two sectors become strongly coupled in the infrared, where the dynamics
can be controlled using holomorphy in the 4-dimensional description.
The two resulting non-perturbative gaugino condensates were shown to provide a
stabilizing potential for the radius modulus.
In order to cancel the effective 4-dimensional cosmological constant 
a source of
spontaneous supersymmetry breaking is required.
We analyzed the simplest possibility that a supersymmetry-breaking
sector is also localized on the hidden brane.

The stabilized radius is in the regime where the warp factor effect is large
provided that $(i)$ the hidden brane gaugino condensate is small 
compared to the
5-dimensional Planck scale;
and $(ii)$ the bulk radius of curvature $1/k$ is not much larger
than the bulk super-Yang--Mills coupling $g_5^2$.
Neither condition requires any fine-tuning.
In particular, the first condition is very natural, since the non-perturbative
gaugino condensate is exponentially suppressed in terms of the 
fundamental gauge
coupling.

We emphasize that the fact that the radius potential is dominated by
non-pertur\-bative super-Yang--Mills dynamics is crucially dependent on
supersymmetry.
In a non-supersymmetric theory, there would be perturbative 
contributions to the
radius potential at the compactification scale from Casimir energy that would
dominate the exponentially smaller contribution from non-perturbative bulk
Yang--Mills dynamics.
In our model, these effects are absent because supersymmetry is unbroken
at the compactification scale.

A heuristic understanding of how stabilization is achieved in our model is to
note that the infrared confinement of the bulk Yang-Mills theory gives a
field-theoretic realization of composite extended states in the bulk, 
namely the
confined hadrons.
The spectrum of such extended states is certainly sensitive to the 
radius and it
is not surprising that their virtual effects can generate a radius
potential.
It is indeed possible that the stabilization role could instead be played by
{\it fundamental} extended objects, in a string/M-theory description.
A virtue of our mechanism is that it involves only the infrared dynamics
of point particles, and is therefore under full theoretical control.

We hope to use the stabilization mechanism presented in this paper as the
basis for further studies of supersymmetric and supersymmetry-breaking
physics in warped compactifications.

\section*{Acknowledgements}
We are very grateful to J. Bagger for discussions of closely related
work prior to publication.
We also thank H. Nishino for discussions.
M.A.L. was supported by NSF grant PHY-98-02551.

\appendix{Appendix A: Derivation of Effective Theory}
In this appendix, we give a complete and self-contained derivation of the
4-dimensional effective lagrangian.

\subsection{\label{genmatch}Matching and Heavy Tadpoles}
We begin by explaining the formalism we will use to integrate out
heavy fields at tree level.
We consider a general classical and local theory of some light fields
$L(x)$ interacting with some heavy fields $H(x)$.
We  will truncate the effective lagrangian at two-derivative order,
higher derivatives  being subdominant at long wavelengths.
While $x$ denotes a point in a spacetime of fixed dimensionality (4 in the
case of interest) this spacetime need not be exactly flat but may have small
curvatures relative to the heavy masses.
We will be interested in the case where the heavy fields are
an infinite tower of KK states;
however we will suppress indices
on the fields since it will be obvious where they go at the end.

Let $S[H, L]$ denote a local classical action that we start with.
We will assume (by shifting the definitions of fields if necessary) that $L = H
= 0$ is a classical solution, and we will expand our theory about this `vacuum'
solution.
Expanding the action in heavy fields and $x$
derivatives gives 
\beq\bal
S[H, L] = S_{\rm light}[L] + \myint d^4 x \Bigl[ &
\lambda(L) H - \sfrac{1}{2} M^2(L) H^2 +
\Phi(L) (\partial L) H
\\
&+ \scr{O}(\partial^2 H) +
    {\cal O}(H \partial H) ~+~ {\cal O}(H^3) \Bigr].
\eal\eeq
$S_{\rm light}[L]$ consists of the part of the fundamental action which is
independent of $H$;
by assumption the mass terms in $S_{\rm light}$ are small compared to
$M^2(L = 0)$,
the mass scale of the heavy fields.
Note that the remaining terms in the action contain terms linear in $H$,
which we call `heavy tadpole' terms.
The first two terms in the integral contain all terms linear and quadratic in
$H$ but containing no derivatives.
The third term contains all terms linear in $H$ with at least one derivative,
which by integration by parts can be taken to act only on light fields.
The remaining terms contain terms linear in $H$ with two or more derivatives,
terms quadratic in $H$ with one or more derivative, and terms of cubic and
higher order in $H$.

Without loss of generality we can set $\lambda \equiv 0$, by making the field
redefinition
\beq
H \rightarrow H + \frac{\lambda}{M^2(L)}.
\eeq
Since the fields $H$ are heavy by assumption we can expand this in powers
of $L$, with higher-order terms suppressed by $M^2(L = 0)$, the mass scale
of the heavy fields.
With this choice, the only heavy tadpoles involve derivatives.

The equations of motion for $H$ then read,
\beq[Hsoln]
    H = \frac{1}{M^2(L)} \left[ \Phi(L) \partial L +
{\cal O}(\partial^2) + {\cal O}(\partial H) +  {\cal O}(H \partial L)
+ {\cal O}(H^2) \right].
\eeq
(In the $\scr{O}(\partial^2)$ terms, the derivatives act on light fields.)
This equation can be solved iteratively by expanding in powers of $L$,
starting with the leading order solution
\beq[Hsolnlead]
H = \frac{\Phi(L) \partial L}{M^2(L)}
+ \scr{O}(\partial^2).
\eeq
Subleading terms are suppressed by additional powers of $M^2(L)$.

We now substitute the solution for $H$
back into the fundamental
action, thereby obtaining an effective action purely for the light fields.
To determine the long-wavelength action up to two derivative order,
only the leading order solution \Eq{Hsolnlead} for $H$ is required.
At this order, we therefore obtain
\beq
S_{\rm eff}[L] = S_{\rm light}[L] + \myint d^4 x\,
\frac{ \left[ \Phi(L) \partial L \right]^2}{2 M^2(L)} +
{\cal O}(\partial^3) .
\eeq
We see that at 2-derivative order
there is a correction to the na\"\i{}ve effective action $S_{\rm light}[L]$
when the original action has heavy tadpoles with one derivative.

\subsection{Radion Effective Field Theory}
We now apply the ideas above to derive the effective lagrangian for the
radion effective lagrangian.
We parameterize the light modes by generalizing the
solution for the metric \Eq{RSsoln} by
$r_0 \to r(x)$, $\eta_{\mu\nu} \to g_{\mu\nu}(x)$:
\beq[radansatz]
ds^2 = e^{-2k r(x) |\ths|} g_{\mu\nu}(x) dx^\mu dx^\nu
+ r^2(x) d\ths^2.
\eeq
Note that $g_{\mu\nu}(x)$ transforms under 4-dimensional general coordinate
transformations as a 2-index tensor, and therefore its couplings in the
4-dimensional action are precisely those of the 4-dimensional metric.
There are no non-derivative couplings of $g_{\mu\nu}$ provided we
cancel the 4-dimensional cosmological constant.
Also note that $r(x)$ is derivatively coupled, since $r(x) = r_0$ is a solution
for any constant $r_0$.
Therefore the action $S_{\rm light}$ obtained by substituting the
metric \Eq{radansatz} into the 5-dimensional action does not contain
mass terms for the light fields.

We parameterize the heavy modes in terms of the 5-dimensional metric
\beq[heavyparam]
\bal
ds^2 &= e^{-2k r(x) |\ths|}
\left[ g_{\mu\nu}(x) + H_{\mu\nu}(x, \ths) \right] dx^\mu dx^\nu
\\
&\qquad
+ 2 H_{\ths\mu}(x, \ths) d\ths dx^\mu
+ r^2(x) \left[ 1 + H_{\ths\ths}(x, \ths) \right] d\ths^2.
\eal\eeq
This must be supplemented with a restriction on $H_{\mu\nu}$ to ensure
that it is `orthogonal' to the zero mode $g_{\mu\nu}$, and we must impose
a gauge on the fluctuations $H_{MN}$.
The details of this will not be needed for our discussion.

As explained in Section~\ref{genmatch}, the correct effective action
at 2-derivative order differs
from $S_{\rm light}$ if there are heavy tadpoles containing a single
$x$ derivative.
By 4-dimensional Lorentz invariance, the only terms of this form involve
the metric fluctuation $H_{\ths\mu}$, \eg $\partial^\mu r H_{\ths\mu}$.
Direct calculation shows that this vanishes in the metric \Eq{heavyparam}.
Therefore, there are no corrections to the effective action at 2-derivative
order, and the correct effective action is obtained simply by using the
metric \Eq{radansatz}.
This gives
\beq[leff4]
S_{\rm 4,eff} = -\frac{M_5^3}{k} \myint d^4 x\, \sqrt{-g} \left[
\left( 1 - e^{-2 \pi k r(x)} \right) R(g)
+ \cdots \right]
\eeq

\subsection{The Graviphoton}
We now turn to the graviphoton $B_M$.
In this case, there is a classical solution $B_\mu \equiv 0$,
$B_\ths \equiv b_0$ for constant $b_0$.
In analogy with the radion, we parameterize the light modes by
generalizing this solution by $b_0 \to b(x)$:
\beq[bdef]
B_\mu \equiv 0,
\qquad
B_\ths(x, \ths) = b(x).
\eeq
In this case there are $\scr{O}(\partial_\mu)$ heavy tadpoles involving the
massive modes $B_\mu$:
\beq[Bmutad]
S_5 = - M_5^3 \myint d^5 X\,
\partial_\ths \left[ \sqrt{-G}\, G^{\mu\nu} G^{\ths\ths} \partial_\nu B_{\ths}
\right] B_\mu + \scr{O}(B_\mu^2) + \scr{O}(\partial_\mu^2).
\eeq
Here $G_{MN}$ is the 5-dimensional metric \Eq{radansatz}, which includes
the light modes $g_{\mu\nu}(x)$ and $r(x)$.

As explained in Section~\ref{genmatch}, the presence of the tadpole \Eq{Bmutad}
means that there are corrections to the effective lagrangian at
$\scr{O}(\partial_\mu^2)$.
We must therefore integrate out the heavy fields $B_\mu$,
including the effects of the tadpole in \Eq{Bmutad}.
The fields $B_\mu$ have nonzero KK masses because they are odd under
the orbifold ${\bf Z}_2$; the mass terms are contained in the
$\scr{O}(\partial_\ths^2 B_\mu^2)$ terms in the action.
Including these mass terms and the $B_\mu$ tadpole in \Eq{Bmutad}, the $B_\mu$
equation of motion is
\beq
\partial_\ths \left[ e^{-2k |\ths| r(x)} \left(
\partial_\ths B_\mu(x, \ths) - \partial_\mu b(x) \right) \right] = 0.
\eeq
The solution is
\beq
e^{-2k |\ths| r(x)} \left(
\partial_\ths B_\mu(x, \ths) - \partial_\mu b(x) \right)
= c_\mu(x)
\eeq
where $c_\mu(x)$ is independent of $\ths$.
The function $c_\mu(x)$ is determined by demanding the periodicity of
$B_\mu$ in $\ths$:
\beq
c_\mu(x) = -2 \pi k r(x) \frac{\partial_\mu b(x)}{e^{2 \pi k r(x)} - 1}.
\eeq
We now substitute this back into the action using the result for the
graviphoton field strength
\beq
C_{\ths\mu}(x, \ths) = e^{+2 k |\ths| r(x)} c_\mu(x).
\eeq
In this way, we obtain
\beq[db2app]
\De S_{\rm 4,eff} = -2 \pi^2 M_5^3 k \myint d^4 x \sqrt{-g}\,
\frac{\partial^\mu b \partial_\mu b}{e^{2 \pi k r} - 1}.
\eeq

\subsection{The Radion Supermultiplet}
We have derived the low-energy effective theory in terms of $r(x)$ and
$g_{\mu\nu}(x)$ (defined by \Eq{radansatz}) and $b(x)$ (defined by \Eq{bdef}).
In a manifestly supersymmetric description, these degrees of freedom can be
parameterized by a \N supergravity multiplet and a chiral superfield $T$.
We wish to find the relation between the fields $g_{\mu\nu}(x)$, $r(x)$,
and $b(x)$, and the components of the supermultiplets in an off-shell
supersymmetric formulation.
To do this it is useful to couple the 5-dimensional theory
to various probes, and track how these probes appear in the 4-dimensional
effective action.
Matching the component and manifestly supersymmetric forms of the
4-dimensional action gives the relation between the component fields
and superfields.

We first couple the SUGRA theory to a bulk SYM multiplet.
The additional massless bosonic
fields in the 4-dimensional effective theory are then the gauge field $A_\mu$
and an adjoint scalar $\Phi$.
Because these both transform in the adjoint representation of the gauge
group, there is no possibility of mixing between the gauge and gravitational
modes in the 4-dimensional effective theory.
The SYM zero modes form a 4-dimensional \N SYM multiplet.
The vector zero mode is given by
\beq[SYMzeromodeapp]
A_\mu(x, \ths) = a_\mu(x).
\eeq
The fact that the zero mode is independent of the warp factor is due to
the classical conformal invariance of Yang--Mills theory.

In the 4-dimensional theory effective theory, the gauge kinetic
term can be written in the manifestly supersymmetric form
\beq
\De \scr{L}_{\rm 4,eff} = \myint d^2\th\, S(T) \tr(W^\al W_\al) + \hc,
\eeq
where $S(T)$ is holomorphic.
We will make a holomorphic field redefinition $S(T) \to T / g_5^2$
so that the action becomes
\beq[Tdef]
\De \scr{L}_{\rm 4,eff} = \myint d^2\th\, \frac{T}{g_5^2} \tr(W^\al 
W_\al) + \hc
\eeq
Expanding this in components, we see that
\beq
\frac{T}{g_5^2} = \frac{1}{2 g_4^2} + \frac{i\Th}{16\pi^2} + \cdots
\eeq
where $g_4$ is the gauge coupling and $\Th$ is the gauge theta angle.
Substituting \Eqs{radansatz} and \eq{SYMzeromodeapp}
into the 5-dimensional SYM action and integrating
over $\ths$ gives
\beq
\frac{1}{g_4^2} = \frac{2\pi r}{g_5^2},
\eeq
which yields
\beq[ReTapp]
\Re(T) = \pi r.
\eeq
The gauge theta angle gets a contribution from the graviphoton from the
5-dimen\-sional SUGRA coupling \cite{gun}
\beq
\De\scr{L}_5 = -\frac{1}{2\sqrt{6}\, g_5^2}
\ep^{MNPQR} B_M \tr F_{NP} F_{QR},
\eeq
which gives
\beq
\frac{\Th}{16\pi^2} = \frac{2 \pi}{\sqrt{6}}\, b.
\eeq
We therefore obtain
\beq[ImTapp]
\Im(T) = \frac{2 \pi}{\sqrt{6}}\, b.
\eeq


\subsection{Supersymmetry and Weyl Rescaling}
At two derivative order, the most general locally \N supersymmetric
lagrangian for the radion chiral multiplet $T$ can be written
\beq[SUGRA4app]
\scr{L}_{\rm SUGRA,4}
&= \myint d^4 \th\, \phi^\dagger \phi f(T, T^\dagger)
\\
&= \sqrt{-\bar{g}} \biggl[
-\frac{1}{6} f R(\bar{g})
- \frac{1}{4 f} \bar{g}^{\mu\nu}
(f_T \partial_\mu T - \hc) (f_T \partial_\nu T - \hc)
\nonumber\\
\eql{gencompLapp}
& \qquad\qquad
- f_{T^\dagger T} \bar{g}^{\mu\nu} \partial_\mu T^\dagger \partial_\nu T
+ \hbox{\rm fermion\ terms} \biggr].
\eeq
Note that the metric $\bar{g}_{\mu\nu}$ that appears here is not assumed to be
the same as the metric $g_{\mu\nu}$ introduced above.
The most general relation between them compatible with general coordinate
invariance is%
\footnote{By 4-dimensional parity, $h$ must be an even function of $b$.}
\beq
\bar{g}_{\mu\nu} = h(r, b) g_{\mu\nu}.
\eeq
The function $h$ is not well-defined until we completely fix the definition of
$\bar{g}_{\mu\nu}$ in the manifestly supersymmetric theory.
We do this by considering a probe consisting of a superpotential term
$\int d^2 \th\, J$ localized on the hidden brane at $\ths = 0$.
In the 4-dimensional effective theory, this gives rise to
\beq
\De\scr{L}_{\rm 4,eff} = \myint d^2\th\, \phi^3 \, \ell(T) J + \hc
\eeq
where $\phi$ is the conformal compensator and $\ell(T)$ is holomorphic.

We can now make a field redefinition
to set $\ell(T) \equiv 1$.
This can be done by means of a \Kahler transformation.
In the superconformal formalism, this is a redefinition of the
conformal compensator
\beq
\phi' = \left[ \ell(T) \right]^{1/3} \phi
\eeq
that is made \emph{prior} to fixing the superconformal gauge \Eq{scgauge}.
That is, we break the superconformal invariance by the choice
\beq
\phi' = 1 + \th^2 F'_{\phi}.
\eeq
Note that since the gauge kinetic term is classically scale invariant,
it is independent of $\phi$.
Therefore this does not affect the field definition made in \Eq{Tdef}.
In components, this field redefinition involves a Weyl rescaling of the
metric $\bar{g}_{\mu\nu}$, and fixes its definition completely.

With this choice, we now compare the effective action for the brane
superpotential to what is obtained by substituting the metric \Eq{radansatz}
into the component form.
In the supersymmetric form, the brane action is independent of $T$, and in
the component form it is independent of $r$, $b$.
This can only be the case if
\beq
\bar{g}_{\mu\nu} = g_{\mu\nu}.
\eeq

Having established this, we can read off the function $f$
from \Eq{leff4}.
Note that the graviphoton Aharonov--Bohm phase does not contribute to the
coefficient of $R(g)$ in the effective action.
Therefore,
\beq[fapp]
f = \frac{3 M_5^3}{k} \left( e^{-\pi k (T + T^\dagger)} - 1 \right).
\eeq
Having determined $f$, the remaining terms in \Eq{gencompLapp} are fixed.
With the identification of $T$ in \Eqs{ReTapp} and \eq{ImTapp}, we have
checked that these terms agree with the direct component calculation
of the $(\partial r)^2$ and $(\partial b)^2$ terms.
In particular, both the nontrivial functional form and the
coefficient of the graviphoton kinetic term \Eq{db2app} agree with
\Eq{gencompLapp} with $f$ given by \Eq{fapp}.

\subsection{Brane Couplings}
We now consider  arbitrary couplings localized on the hidden brane.
In the 4-dimen\-sional effective theory, local \N SUSY implies that these
take the form
\beq
\bal
\scr{L}_{\rm 4,hid} &= \myint d^4 \th\, \phi^\dagger \phi\,
f_{\rm hid}(\Si, \Si^\dagger, T, T^\dagger)
\\
& \qquad + \myint d^2 \th \left[
S_{\rm hid}(\Si, T) \tr(W'^\al W'_\al) + \phi^3 {\cal W}_{\rm hid}(\Si, 
T) \right]
+ \hc
\eal\eeq
In the coordinates we have chosen, the induced metric on the hidden brane
is independent of $r$ (see \Eq{radansatz}).
Therefore, by locality $\scr{L}_{\rm 4,hid}$ is independent of $r$.
Since $S_{\rm hid}$ and ${\cal W}_{\rm hid}$ are holomorphic, this immediately
implies that they are independent of $T$.
For the non-holomorphic function $f_{\rm hid}$, the argument requires a
few steps.
Because there are no derivative couplings involving $r$, we have
\beq
f_{\rm hid} = c \cdot (T - T^\dagger)
+ \hbox{\rm independent\ of\ } T,
\eeq
where $c$ is a constant.
Because the 4-dimensional Planck scale is independent of the
Aharonov-Bohm phase $b_0$, we have $c = 0$.
Therefore, $f_{\rm hid}$ is also independent of $T$, and we have
\beq[Lhidapp]
\bal
\scr{L}_{\rm 4,hid} &= \myint d^4 \th\, \phi^\dagger \phi\,
f_{\rm hid}(\Si, \Si^\dagger)
\\
& \qquad + \myint d^2 \th \left[
S_{\rm hid}(\Si) \tr(W'^\al W'_\al) + \phi^3 {\cal W}_{\rm hid}(\Si) \right]
+ \hc
\eal\eeq

For couplings localized on the visible brane, the induced metric is
$e^{-2 \pi k r(x)} g_{\mu\nu}(x)$, and the couplings localized on
the visible brane will depend on $T$.
Using arguments similar to those above, it is easy to see that the
result is
\beq[Lvisapp]
\bal
\scr{L}_{\rm 4,vis} &= \myint d^4\th\, \phi^\dagger \phi\,
e^{-k(T + T^\dagger)}
f_{\rm vis}(Q, Q^\dagger) 
\\
&\qquad + \myint d^2\th \left[ S_{\rm vis}(Q) \tr \tilde{W}^\al \tilde{W}_\al
+ \phi^3 e^{-3 k T} {\cal W}_{\rm vis}(Q) \right] 
+ \hc,
\eal\eeq
Summarizing, the full 4-dimensional effective lagrangian is the sum of
\Eqs{SUGRA4app}, \eq{Lhidapp}, and \eq{Lvisapp}.


\end{document}